\def\lesssim{\mathrel{\hbox{\rlap{\hbox{\lower4pt\hbox{$\sim$}}}\hbox{$<$}}}}

\def\gtrsim{\mathrel{\hbox{\rlap{\hbox{\lower4pt\hbox{$\sim$}}}\hbox{$>$}}}}

\def\msun{$M_{\odot}$}

\def\teff{$T_{\rm eff}$~}

\def\ll_lsun{log$({L/\rm L_{\odot}})$~}

\def\masa_msun{$M/ \rm M_{\odot}$~}

\def\m_mstar{$M/M_{*}$~}

\documentclass{aa}
\usepackage{graphicx}
        
\begin{document}

\title{The double-layered chemical structure in DB white dwarfs}

\author{L. G. Althaus\thanks{Member of the Carrera del Investigador
Cient\'{\i}fico y Tecnol\'ogico, CONICET, Argentina.} and
A. H. C\'orsico\thanks{Fellow of the Consejo Nacional de 
Investigaciones Cient\'{\i}ficas y T\'ecnicas (CONICET), Argentina.}}
\offprints{L. G. Althaus}

\institute{Facultad  de Ciencias  Astron\'omicas  y Geof\'{\i}sicas,  
Universidad Nacional  de  La  Plata,  Paseo  del  Bosque  S/N,  (1900)  
La  Plata, Argentina.\\ Instituto de Astrof\'{\i}sica La Plata, IALP, 
CONICET\\ \email{althaus,acorsico@fcaglp.unlp.edu.ar}}

\date{Received; accepted}
                
\abstract{The purpose of this work is to study the structure and
evolution of white dwarf stars  with helium-rich atmospheres (DB) in a
self-consistent  way with  the predictions  of  time-dependent element
diffusion. Specifically,  we have  considered white dwarf  models with
stellar masses in the  range 0.60-0.85\msun\ and helium envelopes with
masses from  $10^{-2}$ to $10^{-4} M_*$.  Our  treatment of diffusion,
appropriate for multicomponent  gases, includes gravitational settling
and  chemical and  thermal  diffusion.  OPAL  radiative opacities  for
arbitrary  metallicity  and  carbon-and oxygen-rich  compositions  are
employed.    Emphasis   is   placed    on   the   evolution   of   the
diffusion-modeled  double-layered chemical structure.  This structure,
which is characterized by a  pure helium envelope atop an intermediate
remnant  shell rich  in helium,  carbon  and oxygen,  is expected  for
pulsating  DB white  dwarfs,  assuming that  they  are descendants  of
hydrogen-deficient PG1159 post-AGB stars.   We find that, depending on
the  stellar mass, if  DB white  dwarf progenitors  are formed  with a
helium content  smaller than  $\approx 10^{-3} M_*$,  a single-layered
configuration is  expected to emerge during the  DB pulsation 
instability strip.
We  also explore  the  consequences of  diffusively evolving  chemical
stratifications  on the  adiabatic  pulsational properties  of our  DB
white dwarf models.  In this context, we find  that the evolving shape
of the  chemical profile translates  into a distinct behaviour  of the
theoretical period distribution as compared with the case in which the
shape  of the  profile is  assumed to  be fixed  during  the evolution
across  the  instability  strip.   In  particular, we  find  that  the
presence  of  a double-layered  structure  causes  the period  spacing
diagrams to exhibit mode-trapping substructures. Finally, we extend 
the scope of the
calculations to  the domain of the  helium-rich carbon-contaminated DQ
white dwarfs.  In particular, we  speculate that DQ white  dwarfs with
low detected carbon abundances would  not be descendants of the PG1159
stars.
 }  

\authorrunning{Althaus and C\'orsico}

\titlerunning{The double-layered chemical structure in DB white dwarfs}

\maketitle


\section{Introduction}

White dwarf  stars with helium-rich atmospheres  (DB) constitute about
the 20  \% of all observed  white dwarfs. The majority  of these stars
are thought to be the result  of a very late thermal pulse experienced
by  post-asymptotic-giant-branch (post-AGB)  progenitors  during their
early cooling phase  (the born-again scenario; see Iben  et al. 1983).
As a result of such a pulse, most of the residual hydrogen envelope is
completely burnt,  and the  star returns to  the red giant  region and
then  to the  planetary nebula  regime at  high  effective temperature
(\teff) as  a helium-burning object.   Evolutionary calculations which
incorporate diffusive overshooting (Herwig et al.  1999) show that the
occurrence  of a very  late thermal  pulse leads  to the  formation of
hydrogen-deficient post-AGB  stars with surface  abundances of helium,
carbon  and oxygen.   This prediction  is in  good agreement  with the
surface  abundance pattern observed in most PG1159 stars  or  the WR-type
central stars of  planetary nebulae (Dreizler \& Heber  1998), the hot
and hydrogen-deficient  remnants of post-AGB  stars generally believed
to be the immediate progenitors of most DB white dwarfs.
 
The  detection of  pulsations  in  some DB  white  dwarfs 
(in the temperature range of $\approx 27000-20000$K) has  allowed
researchers  to  provide   independent valuable  constraints  on  the
internal structure and  evolution of these stars, as  well as on their
progenitors.   Indeed, asteroseismological  techniques  have become  a
powerful  tool for  inferring  fundamental properties  such the  white
dwarf mass,  effective temperature, core composition  and helium layer
mass (see Bradley \& Winget 1994; Metcalfe et al. 2000 and Metcalfe et
al.  2001). In particular, Metcalfe
et al. 2001  (see also Metcalfe et al. 2002)  have recently applied DB
white   dwarf   asteroseismology   to   place   constraints   on   the
$^{12}$C($\alpha,\gamma)^{16}$O  reaction rate  from inferences  for the
abundance of central oxygen in the pulsating DB GD 358. It is worth noting
that such studies are based on stellar models for pulsating DBs with
a pure helium envelope atop the carbon-oxygen core (single-layered
envelope).

However, recent theoretical evidence  makes past inferences about core
composition of DBs and the $^{12}$C($\alpha,\gamma)^{16}$O reaction rate
somewhat  uncertain. Indeed,  Fontaine \&  Brassard (2002)  have shown
that   the  diffusion-built-up,   double-layered   chemical  structure
expected for most of the pulsating DBs leads to a distinct theoretical
period distribution from  that predicted by stellar models  in which a
single-layered   configuration  is   assumed.   The   presence   of  a
double-layered  envelope  in  DB   white  dwarfs  has  been  found  in
evolutionary  calculations including time-dependent  element diffusion
by Dehner \&  Kawaler (1995) and Gautschy \&  Althaus (2002), assuming
that  DB  white dwarfs  descend  from  PG  1159 stars.   Indeed,  such
calculations show that  by the time the domain of  the variable DBs is
reached, models are characterized by two different chemical transition
zones (double-layered configuration).  In fact, above the carbon-oxygen
core, there exists an envelope consisting of an intershell region rich
in helium,  carbon and  oxygen, the relics  of the  short-lived mixing
episode  occurred  during  the  last  helium  thermal  pulse,  and  an
overlying pure helium mantle  resulting from gravitational settling of
carbon   and  oxygen.    Finally,   DB  asteroseismological   fittings
incorporating both  the double-layered envelope  feature expected from
time-dependent  diffusion  calculations  and adjustable  carbon-oxygen
cores have  recently been  presented by Metcalfe  et al. (2003)  for a
wide range  of helium contents  and stellar masses. Despite  their models
yielding significantly  better fits to observations,  the derived stellar
parameters   for   some  fittings   led   them   to  conclude   that
double-layered models  with adjustable carbon/oxygen cores may not be 
entirely appropriate  to explain the observations.

In  view of  these  concerns, this  paper  is aimed  at exploring  the
evolution  of  DB white  dwarfs  in  a  self-consistent way  with  the
predictions of time-dependent element diffusion. Emphasis is placed on
the chemistry  variations along  the evolutionary stages  during which
these stars are expected to  pulsate. Specifically, we hope to address
the  following question:  Is it  possible that  the diffusion-induced,
double-layered  structure   could  be  altered   by  further  chemical
evolution  to  such  an  extent that  a  single-layered  configuration
emerges before the star reaches  the red edge of the instability strip?. In
view of  the lack of  previous calculations of the  mass-dependence of
time-dependent diffusion  profiles, we judge that this  question is an
important one  to be answered. Additionally, we  extend the scope of  
the paper and explore  the consequences of  diffusively evolving  
chemical stratifications for the adiabatic pulsational properties 
of the models. Finally, the implications of our results
for the carbon surface abundance in some helium-rich DQ white dwarfs 
are explored.

\section{Computational details and input physics}

The  results presented in  this work  have been  obtained with  the DB
white  dwarf  evolutionary code  developed  at  La Plata  Observatory.
Except for  minor modifications, the code is  basically that described
in  Gautschy  \& Althaus  (2002)  and  references  cited therein.   In
particular, microphysics  includes an updated version  of the equation
of state of  Magni \& Mazzitelli (1979), OPAL  radiative opacities for
arbitrary metallicity (Iglesias \&  Rogers 1996) including carbon- and
oxygen-rich  compositions, and up-to-date  neutrino emission  rates and
conductive  opacities  (see  Althaus  et  al.  2002).  In  particular,
opacities for various metallicities  are  required  because of  the
metallicity gradient  that develops  in the envelopes  as a  result of
gravitational  settling and also in the outer layers of DQ white dwarfs 
because of dredge-up episodes (see next section). In this  
work, convection  is treated  in the framework  of   the  mixing  
length   theory  as  given  by   the  ML2 parametrization 
(see Tassoul et al. 1990).

The  evolution  of  the  chemical  abundance  distribution  caused  by
diffusion  processes  has been  fully  accounted  for  by means  of  a
time-dependent,   finite-difference  scheme   that   solves  elemental
continuity  equations.  To  compute  the white  dwarf  evolution in  a
self-consistent way with the  predictions of element diffusion, such a
scheme  is coupled  to  the  evolutionary code.  In  our treatment  of
time-dependent diffusion we have considered gravitational settling and
chemical  and  thermal diffusion  for  the  nuclear species  $^{4}$He,
$^{12}$C  and $^{16}$O.  Radiative levitation,  which is  important in
connection  with surface  composition  of hot  white  dwarfs has  been
neglected.   Specifically, we  have adopted  the treatment  of Burgers
(1969) for multicomponent gases. There, diffusion velocities $w_i$ and
residual heat flows $r_i$ satisfy the set of equations

\begin{eqnarray}
{{{\rm d}p_i} \over {{\rm d}r}}-{{\rho _i} \over \rho}{{{\rm d}p} \over 
{{\rm d}r}}-n_iZ_ieE=
\sum\limits_{j\ne i}^{N} {K_{ij}}\left({w_j-w_i} \right) \cr
+\ \sum\limits_{j\ne i}^{N} {K_{ij}\ z_{ij}} {{m_j\ r_i\ - m_i\
r_j}\over{m_i\ + m_j}},
\end{eqnarray}

\noindent and 

\[{{5} \over {2}}n_i k_{B} \nabla T= - {{5} \over {2}}
\sum\limits_{j\ne
i}^{N} {K_{ij}\ z_{ij}} {{m_j}\over{m_i\ + m_j}}\left({w_j-w_i}
\right)
- {{2} \over {5}}{K_{ii}}\ z_{ii}^{,,}\ r_i \]
\begin{eqnarray}
 -\sum\limits_{j\ne i}^{N} {{K_{ij}}\over{(m_i\ + m_j)^2}}\left(
{3 m_i^2 + m_j^2 z_{ij}^{,} + 0.8m_im_jz_{ij}^{,,}} \right)\
r_i \cr
{+\sum\limits_{j\ne i}^{N} {{K_{ij}m_im_j}\over{(m_i\ +
m_j)^2}}\left(
{3 + z_{ij}^{,} - 0.8z_{ij}^{,,}} \right)\ r_j}.
\end{eqnarray}

\noindent Here, $p_i$,  $\rho_i$, $n_i$,  $Z_i$ and  $m_i$ means,
respectively, the partial pressure, mass density, number density, mean
charge and mass for species $i$ ($N$ means the number of ionic species
plus  electron). $T$,  $k_{B}$  and $\nabla  T$  are the  temperature,
Boltzmann constant and temperature gradient, respectively. The unknown
variables are $w_i$ and  $r_i$  (for  ions  and
electrons). The  electric field  $E$ has also  to be  determined.  The
resistance    coefficients   ($K_{ij},    z_{ij},    z_{ij}^{,}$   and
$z_{ij}^{,,}$)  are from  Paquette at  al. (1986). For more details,
see Gautschy \& Althaus (2002). It is worth mentioning that a recent 
study of the $^{3}$He diffusion in DB white dwarfs has been presented 
by Montgomery et al. (2001). 

\begin{figure}
\centering
\includegraphics[width=250pt]{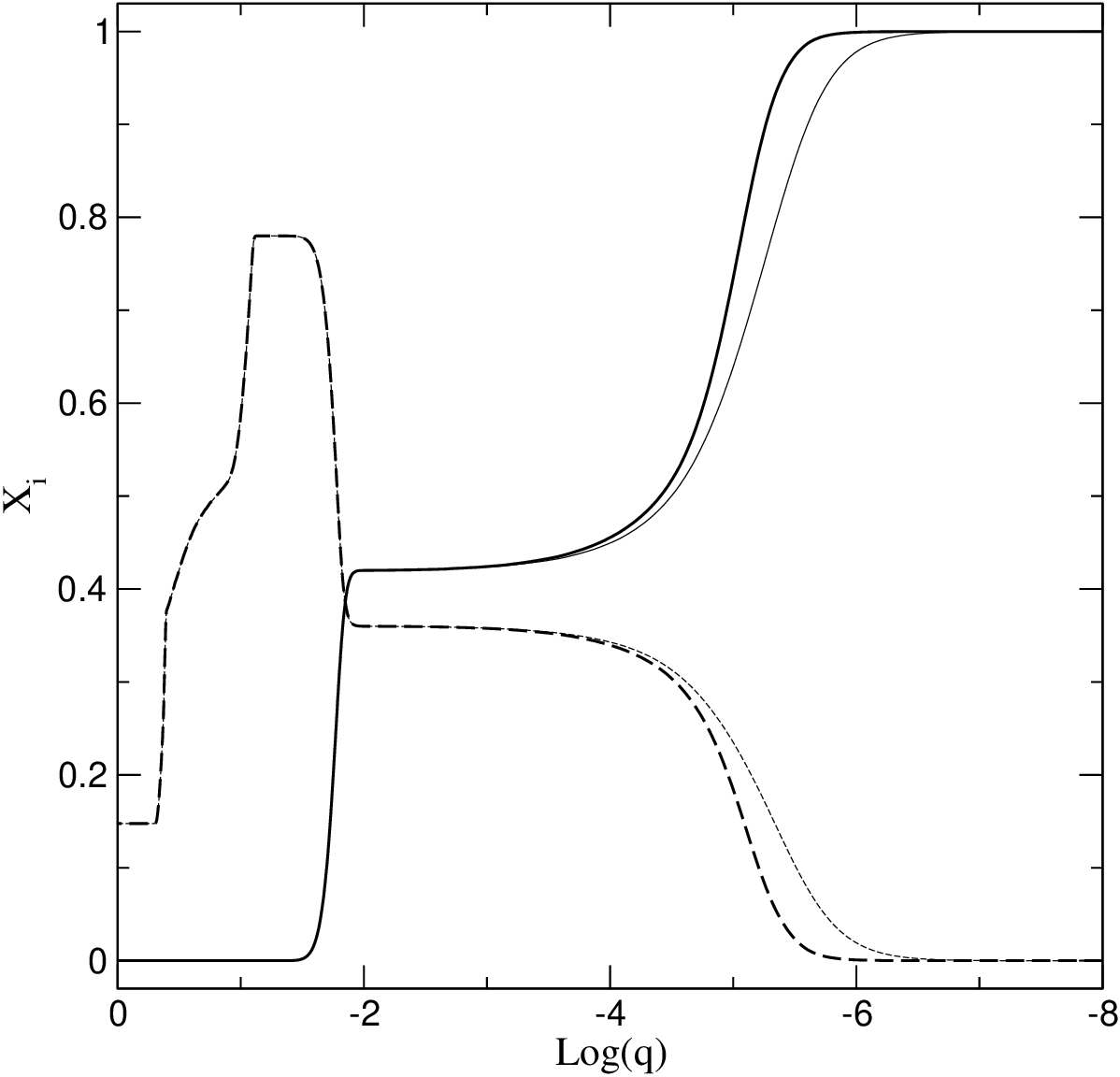}
\caption{Abundance by mass of $^{4}$He and  $^{12}$C (solid and dashed
line, respectively) as a function of the outer mass fraction $q$ for  
0.60-\msun\ DB models at \teff$=27300 K$. Thick (thin) lines
correspond to the situation in which thermal diffusion is considered 
(neglected). Clearly, the inclusion of thermal diffusion leads to 
thicker pure helium mantles.}
\label{termica.eps}
\end{figure}

The shape of the chemical profile in the envelope of a white dwarf
is determined by the competition between
basically partial pressure gradients, gravity and induced electric fields.
In order to get an insight into the essential physics driving element diffusion
and to understand some of the results to be presented later we will
assume for a moment a simplified situation in which only two ionic
species with mean charge $Z_1$ and $Z_2$ and atomic mass number $A_1$ and $A_2$ 
are present. In addition, we assume the electron to have zero mass 
and neglect the contribution from thermal diffusion. It is then straightforward
to show from Eq. (1) that the diffusion velocity satisfies

\[K_{12} w_1 f= \left({{A_2} \over{Z_2}} - {{A_1} \over{Z_1}} \right) 
m_{\rm H} g \ + \]
\begin{eqnarray}
\left({{1} \over{Z_2}} - {{1} \over{Z_1}} \right) k_BT \ {{{\rm d}\ {\rm ln}\ T} 
\over {{\rm d}r}}\ 
+\ {{k_BT} \over{Z_2 n_2}} {{{\rm d} n_2} \over {{\rm d}r}}\ -\ {{k_BT}
\over{Z_1 n_1}} {{{\rm d} n_1} \over {{\rm d}r}},
\end{eqnarray}

\noindent where $w_1$ is the diffusion velocity for specie $1$ ($w_1 < 0$
means that  the element  sinks into the  star) and $f=  (1+A_1 n_1/A_2
n_2) (n_1 Z_1 + n_2 Z_2) / n_1  Z_1 n_2 Z_2 $. In Eq. (3), $m_{\rm H}$ and $g$ are the
hydrogen-atom mass and gravity, respectively. In deriving Eq. (3) we
have assumed  an ideal gas for ions  and use the condition  for no net
mass  flow relative  to the  centre of  mass.  The  first term  on the
right-hand side of Eq. (3)  takes into account the contribution of the
gravitational  settling  and the  influence  of  the induced  electric
field, and the third and  fourth terms refer to the chemical 
diffusion contribution.
We apply Eq. (3) to a  mixture of $^{4}$He and $^{12}$C. Note that the
ionization state of elements  strongly affects the diffusion velocity.
In  particular, in the  non-degenerate outer  layers, $^{12}$C  is not
fully  ionized (for instance, in our \teff$= 27300$ K, 0.6-\msun\ model, 
$^{12}$C becomes completely  ionized at  a mass  depth of  $10^{-8} M_*$  
below  the stellar surface),  and as a result  $^{4}$He diffuses
upwards very rapidly driven mainly by  gravity (first term in Eq. 3 is
dominant).  As cooling  proceeds,  this leads  to  a thickening  pure
helium  mantle.   In  deeper,  more degenerate  layers,  gravitational
settling  turns  out  to  be  less operative.  Indeed,  $^{12}$C  (and
$^{4}$He) becomes fully ionized  and the gravity term vanishes. There,
diffusion is  driven essentially  by chemical gradients,  thus forcing
helium  to penetrate  downwards to  hotter layers.  Note that  this is
quite  different from  the  situation  for a  mixture  of $^{1}$H  and
$^{4}$He.  In that case, the gravity term never vanishes and, at large
degeneracy, it takes over  chemical diffusion, thus halting the inward
diffusion of $^{1}$H  and leading to a complete  separation of the two
species at advanced ages.
    
It is important to note that our diffusion treatment considers thermal
diffusion.   This is  by no  means  a negligible  contribution to  the
diffusion  processes.  Because thermal  diffusion  acts  in the  same
direction as gravitational settling,  the neglect of thermal diffusion
underestimates   the  rate   at  which   carbon  and   oxygen  diffuse
downwards. Hence,  at a  given \teff value,  we expect a  thicker pure
helium  mantle when  thermal  diffusion is  taken  into account.  This
expectation is indeed borne out by Fig. \ref{termica.eps} which 
illustrates the helium
and carbon abundance distribution in  terms of the outer mass fraction
$q$ ($q=1-M_r/M_*$) for  a 0.6-\msun\ DB  white dwarf model at  
\teff$= 27300$ K, i.e.  near the blue edge  of the DB instability  
strip.  Note that
the mass  embraced by  the pure helium  mantle amounts to  $2.5 \times
10^{-6} M_{*}$, as compared with  the value of $6 \times 10^{-7} M_{*}
$ when thermal diffusion is neglected.

With regard to the initial  stellar models needed to start our cooling
sequences,  they  were  obtained  by  means  of  the  same  artificial
procedure as  described in Gautschy \& Althaus  (2002). In particular,
such  starting   models  correspond  to   hydrogen-deficient  post-AGB
configurations  with  an  uniform  envelope  chemical  composition  of
helium, carbon and  oxygen (with abundances of 0.42,  0.36 and 0.22 by
mass, respectively).  As mentioned  in the introduction,  the envelope
composition is the  result of a short-lived mixing event  occurred during the late
thermal pulse  experienced by a  post-AGB remnant.  The  chemical
composition    of    the    core     is    that    of    Salaris    et
al.  (1997). Specifically, we  have considered  DB white  dwarf models
with stellar masses of 0.60, 0.65, 0.70, 0.75 and 0.85 \msun. The mass
of the helium envelope ($M_{\rm He}$)  is constrained by the theory of
post-AGB evolution to be in the range $10^{-2}$ to $10^{-3}$ $M_*$, in
good agreement  with carbon dredge-up simulations in  DQ white dwarfs,
the  cooler descendants  of  DBs, by  MacDonald  et al.  (1998) for  a
typical white  dwarf mass of 0.6  \msun.  In this  work, we considered
the  following  values  for  $M_{\rm He}$:  $8 \times  10^{-3},  9  \times
10^{-4}$ and $ 10^{-4}$ $M_*$.
 
In what follows, we describe the main results of our calculations.  We
want  to  mention   that  this  paper  is  not   aimed  at  performing
asteroseismological  fittings  to  a  particular object.   Rather,  we
concentrate mainly  on the consequences  of element diffusion  for the
chemical profiles expected at the DB instability strip. With regard to
pulsational properties, our  attention is directed exclusively towards
the effects of diffusively evolving chemical profiles on the adiabatic
pulsational properties of the DB white dwarf models.  In this sense, a
detailed description of the evolutionary phases prior to the formation
of DB  white dwarfs, particularly  regarding the chemical  profile for
the  carbon-oxygen  core,  is   not  of  primary  importance  for  our
purposes. As for pulsational calculations, we have computed adiabatic,
nonradial $g$-modes of the DB  white dwarf models with the pulsational
code  employed in  C\'orsico et al. (2001), appropriately  modified to
study  pulsating DB  white dwarfs.   In particular,  the  treatment we
follow to assess the Brunt-V\"ais\"al\"a frequency ($N$) is
that proposed by Brassard et al. (1991).

\section{Evolutionary and pulsational results}

\begin{figure}
\centering
\includegraphics[width=250pt]{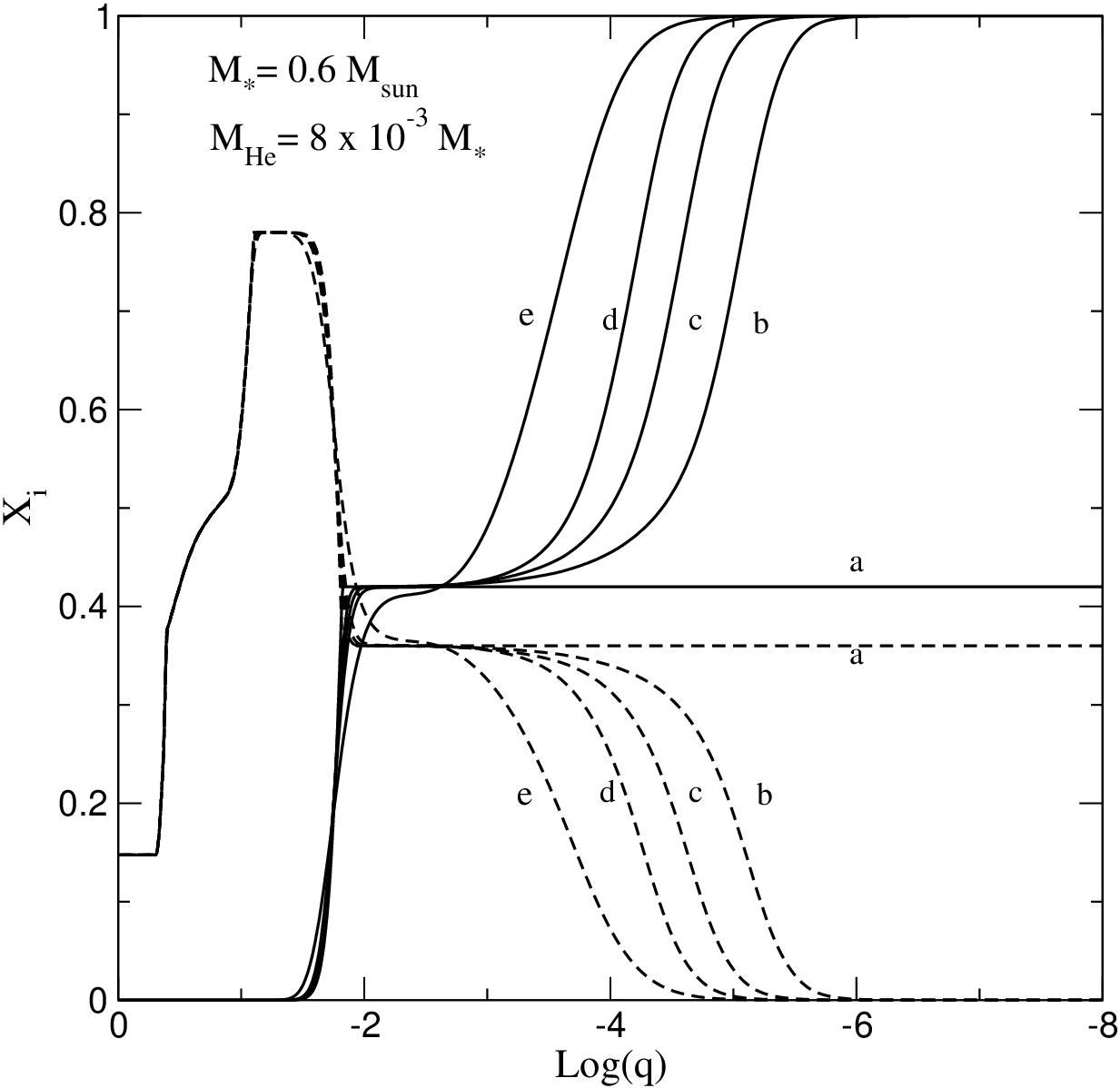}
\caption{Evolution of the abundance by mass of $^{4}$He (solid lines) 
and  $^{12}$C (dashed lines) as a function of the outer mass fraction 
($q$) for 0.60-\msun\ DB models having a helium content of $M_{\rm He}= 
8 \times  10^{-3} M_*$. Starting from a model with an initially
homogeneous envelope (curves a), following models (b, c, d and e)
correspond to evolutionary
stages characterized by $T_{\rm eff}=$ 27500, 23000, 20200 and 10200 K. 
Gravitationally induced diffusion leads to the development of a 
double-layered structure.
Note also that the outer layer abundance distribution
evolves appreciably as the white dwarf cools down through the stages
where DBs are found to pulsate.} 
\label{06_7d-3teff.eps}
\end{figure}

\begin{figure}
\centering
\includegraphics[width=245pt]{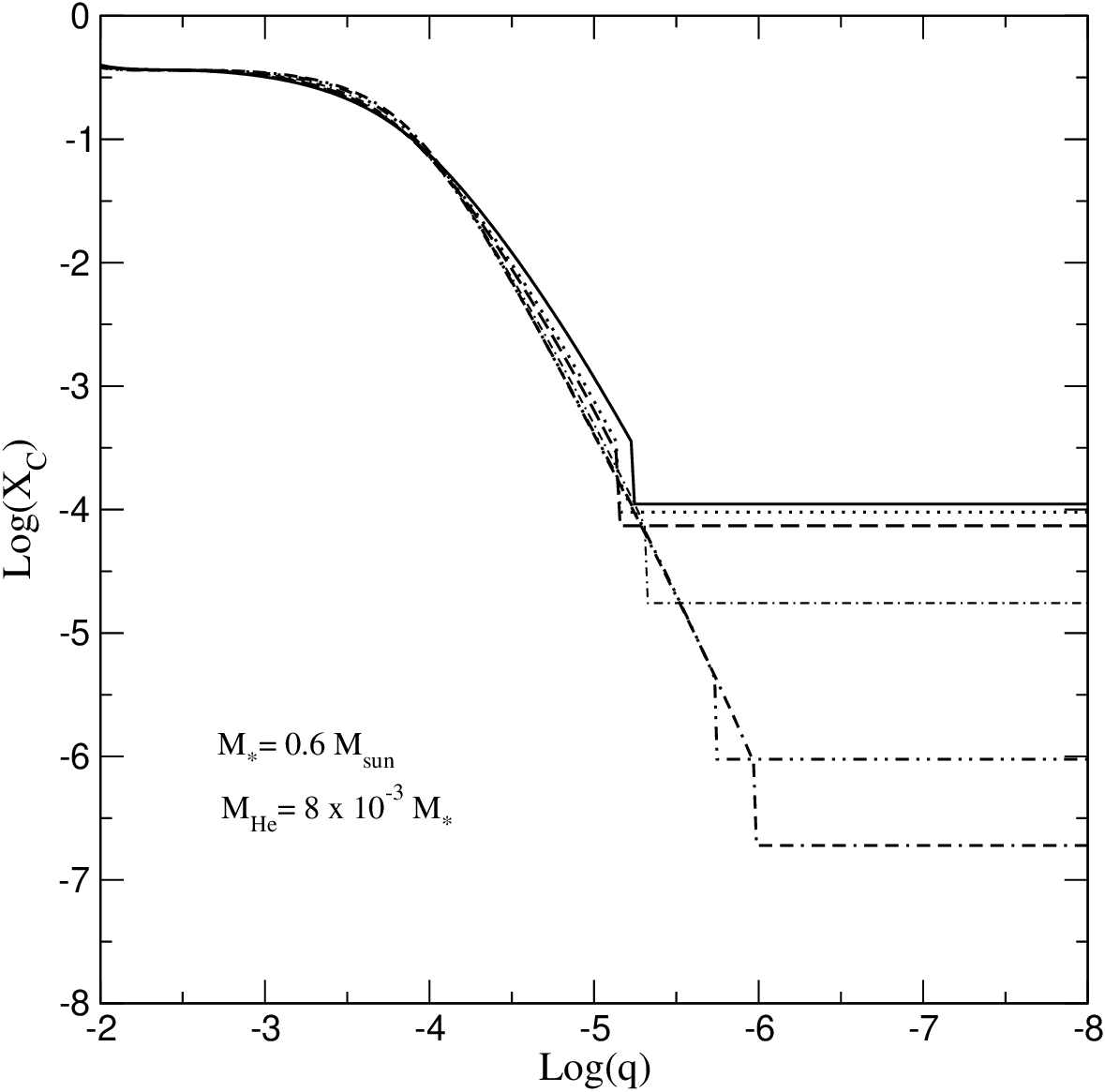}
\caption{Abundance distribution of $^{12}$C  in the outer layers 
at (from bottom to top) $T_{\rm eff}=$ 14000, 13300, 11900, 10700, 
10200 and 9200 K.
Models correspond to a 0.60-\msun\ DB white dwarf having a helium content 
of $M_{\rm He}= 8 \times  10^{-3} M_*$. Note that, as a result of
convection reaching the $^{12}$C diffusive tail, the outer layers become 
substantially $^{12}$C-enriched.} 
\label{dredge.eps}
\end{figure}

We begin by examining Fig. \ref{06_7d-3teff.eps} which illustrates the
evolution  of the  internal $^{4}$He  and $^{12}$C  distribution  as a
function of the outer mass  fraction $q$ for 0.60-\msun\ DB white dwarf
models  with  a  helium  content  of  $8  \times  10^{-3}  M_*$.  In particular,
the expectation  at  three  selected   values  of  \teff  that  cover  the
temperature range for  pulsating DB white dwarfs is depicted.
Clearly, diffusion processes  appreciably modify the chemical profiles
as the white  dwarf evolves, making essentially the  bulk of helium
float  to the  surface  and heavier  element  sink  out of  surface
layers. Note  that, as  found in previous  studies (Dehner  \& Kawaler
1995;  Gautschy \&  Althaus 2002  and Fontaine  \& Brassard  2002) the
formation of  a double-layered structure  is expected by the  time the
white  dwarf  has  reached  the  domain  of  pulsational  instability.
Indeed,  diffusion proceeds  efficiently, giving  rise to  pure helium
outermost  layers.   Specifically, we  find  that  at \teff  $\approx$
27500K, the 0.6-\msun\ DB model is characterized by a pure helium mantle
of  $2 \times  10^{-6} M_*$  (curve  b) and  an underlying  intershell
(still-uniform) region rich in  helium, carbon and oxygen. We note 
that the pure
helium mantles of our models are somewhat larger than those reported by
Dehner  \&   Kawaler  (1995) and  Fontaine   \&  Brassard  (2002). In
part, the difference is due to the inclusion of thermal  diffusion in
our simulations (see section 2). Interestingly,  the
chemical  profile  evolves  appreciably  with  the  further  evolution
through the DB instability strip, as shown by Fig.
\ref{06_7d-3teff.eps}.  It is worth noting that, during such stages,
element diffusion not only thickens the pure helium mantle but also it
modifies the {\it  shape} of the chemical profile,  a feature which is
expected to leave some  traces in the theoretical period distribution,
as compared with the case in which the shape of the profile is assumed
to be fixed during the DB white dwarf evolution across the instability
strip  (see later in this section). 

\begin{figure}
\centering
\includegraphics[width=250pt]{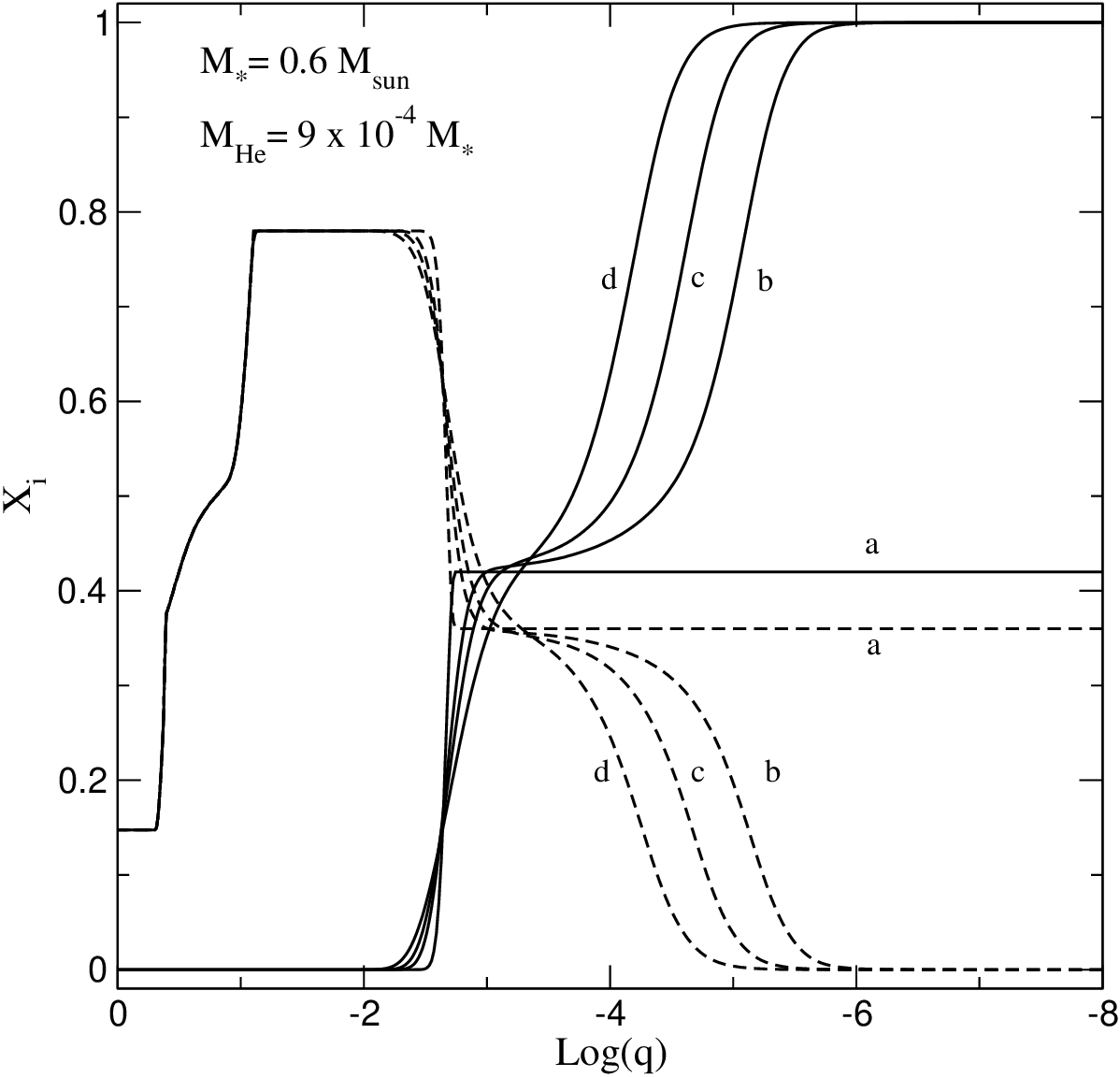}
\caption{Same as Fig. \ref{06_7d-3teff.eps} but for a helium content of 
$M_{\rm He}= 9 \times  10^{-4} M_*$. Starting from a model with an initially
homogeneous envelope (curves a), following models (b, c and d)
correspond to evolutionary stages characterized by $T_{\rm eff}=$ 
27200, 22900 and 19800 K. Here, the shape of the chemical profile is 
notably modified by
diffusion processes. In particular, shortly after the red edge of the 
instability strip is reached, the double-layered structure has virtually
disappeared (curves d).} 
\label{06_9d-4teff.eps}
\end{figure}

Although the  attention of this  paper is focused on  the evolutionary
stage  corresponding to  variable DBs,  we have  followed  the further
evolution of some sequences down  to the domain of the carbon-enriched
DQ  white dwarfs. The  resulting chemical  profile for  the case  of a
0.6-\msun\ DB  model with  a helium content  of $M_{\rm He}=  8 \times
10^{-3}   M_*$   at  \teff=   10200K   is   shown   by  curve   e   in
Fig. \ref{06_7d-3teff.eps}.   Note that even at  such advanced stages,
diffusion has not engulfed the entire helium-carbon-oxygen intershell,
and the  chemical  profile  is  still characterized  by  a  double-layered
structure. In  particular, the  pure helium mantle  extends down  to a
fractional  mass  depth $q  \approx  5  \times10^{-5}M_*$. Note  that,
despite the total  helium content amounting to a  fraction of $\approx
0.01$ of  the stellar  mass, there is  a substantial amount  of carbon
stretching  outwards from  the intershell  region.  We  expect  then a
significant carbon  enrichment in  the surface layers  as a  result of
convective dredge-up of the carbon  diffusive tail, a process which is
generally  believed  to be  responsible  for  the  observed carbon  in
DQs. This  expectation is indeed  borne out by  Fig. \ref{dredge.eps},
which shows  the carbon abundance  distribution in terms of  the outer
mass fraction at various
\teff values for the 0.60-\msun\ DB white dwarf sequence
with a  helium mass of  $8 \times 10^{-3} M_*$. Chemical diffusion causes 
some carbon from deeper layers
to diffuse upwards (see Eq. 3) to where convection dredge it up to the
surface.  As cooling  proceeds, the base of the  convection zone moves
inwards  and eventually  reaches a  maximum depth of 
$4.2 \times10^{-6}$ \msun by  \teff $\approx$
10000K. The depth reached by the convection zone, which at this low effective 
temperature is almost independent of the treatment of convection, is 
limited by the low
conductive opacity of the degenerate interior. It is worth noting
that the presence of carbon
(and dredged metals) in the outer layers renders the helium plasma less transparent,
thus providing a not so deep convection zone.
Our results suggest that  
at \teff= 10000K, the carbon surface
abundance  reaches\footnote{In terms  of a  single-layered  profile, a
pure helium mantle of $\approx 5
\times10^{-5}M_*$ corresponds to a total helium content of
$10^{-3}$-$10^{-4} M_*$. Thus, our  quoted surface carbon abundance is
in  rough comparable  agreement with  the prediction  of  MacDonald et
al.(1998) for their single-layered  sequences with a helium content of
about  $10^{-3} M_*$.}  log $n(^{12}$C$)/n(^4$He$)  \approx  -4$. This
abundance far exceeds the lowest carbon abundance observed in many DQs
(see  MacDonald  et  al.    1998). We speculate that DB models less 
massive than $0.6$ \msun\ evolved directly from PG1159 configurations
should also exhibit such large carbon surface abundances. In fact, even in 
this case, settling of carbon from the outer layers would take long 
time, and accordingly, a substantial amount of carbon would be dredged 
upwards by convection. Additional 
evolutionary calculations for such low-mass DB white dwarfs are  
required to place this statement on a solid background.    
Hence,  the  plausibility  of  an
evolutionary link between low-mass PG1159 and  DQ stars with {\it low} 
detected carbon abundance is not clear if canonical convective dredge-up 
is the source of the observed carbon for such DQs.

\begin{figure}
\centering
\includegraphics[width=250pt]{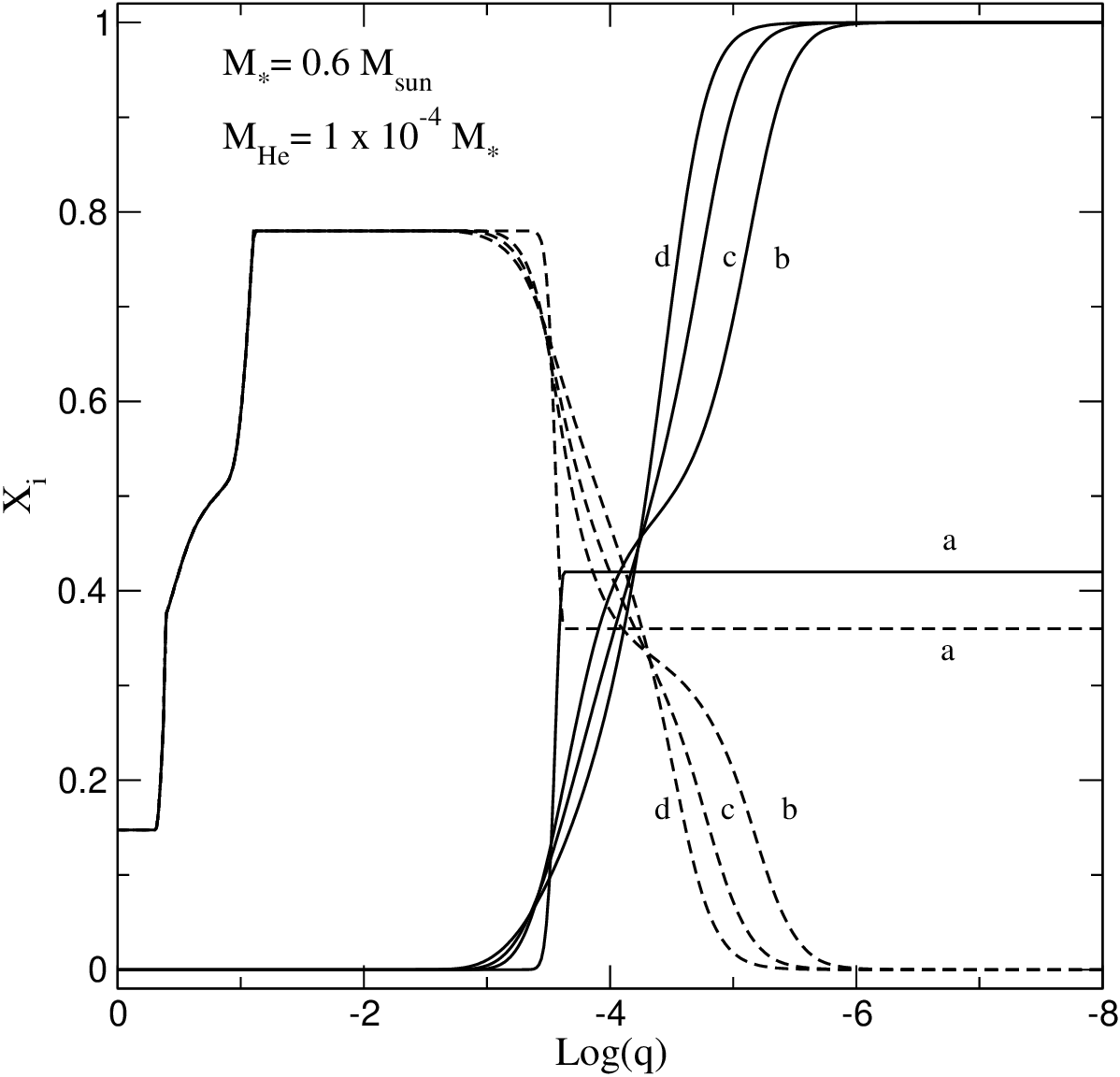}
\caption{Same as Fig. \ref{06_7d-3teff.eps} but for a helium content of 
$M_{\rm He}= 1 \times  10^{-4} M_*$. Starting from a model with an initially
homogeneous envelope (curves a), following models (b, c and d)
correspond to evolutionary
stages characterized by $T_{\rm eff}=$ 27300, 23600 and 20600 K. 
Here, the chemical profile bears no clear signature of a double-layered 
structure even at the hottest \teff values considered (curves b).
Models, instead, are characterized by a single-layered structure.} 
\label{06_1d-4teff.eps}
\end{figure}

In  the case of  small helium  contents, the  shape of the double-layered
profile  is   strongly  affected   by  element  diffusion.  This  is
illustrated by Figs.  \ref{06_9d-4teff.eps} and
\ref{06_1d-4teff.eps}, which show the internal $^{4}$He and $^{12}$C
distribution as a function of $q$ for 0.60-\msun\ DB white dwarf models
with a  helium mass of  $9 \times 10^{-4}  M_*$ and $1  \times 10^{-4}
M_*$, respectively.  On  the basis of these figures,  we conclude that
the double-layered structure is  altered by chemical evolution to such
a degree  that a single-layered configuration emerges  even during the
stages of pulsation instability.  This result is not surprising
in view of Eq. (3). Indeed, for small helium contents, electron degeneracy
occurs below the helium-carbon-oxygen mantle, and as a result, the second term
in Eq (3) is dominant, leading to the formation of a single-layered
structure in relatively short ages.  
Specifically, for $M_{\rm He}= 9
\times 10^{-4} M_*$ a single-layered configuration occurs by the time the 
star reaches the end
of the instability strip, and  for $M_{\rm He}= 1 \times 10^{-4} M_*$,
a single-layered configuration sets in  even by the onset of pulsation
instability. Hence, if post-AGB, DB white dwarf progenitors are formed
with  a  helium  content  smaller  than  $\approx  10^{-3}  M_*$,
the double-layered  structure is
expected to become single-layered by the time the star reaches the red
edge of the DB instability strip.

\begin{figure}
\centering
\includegraphics[width=250pt]{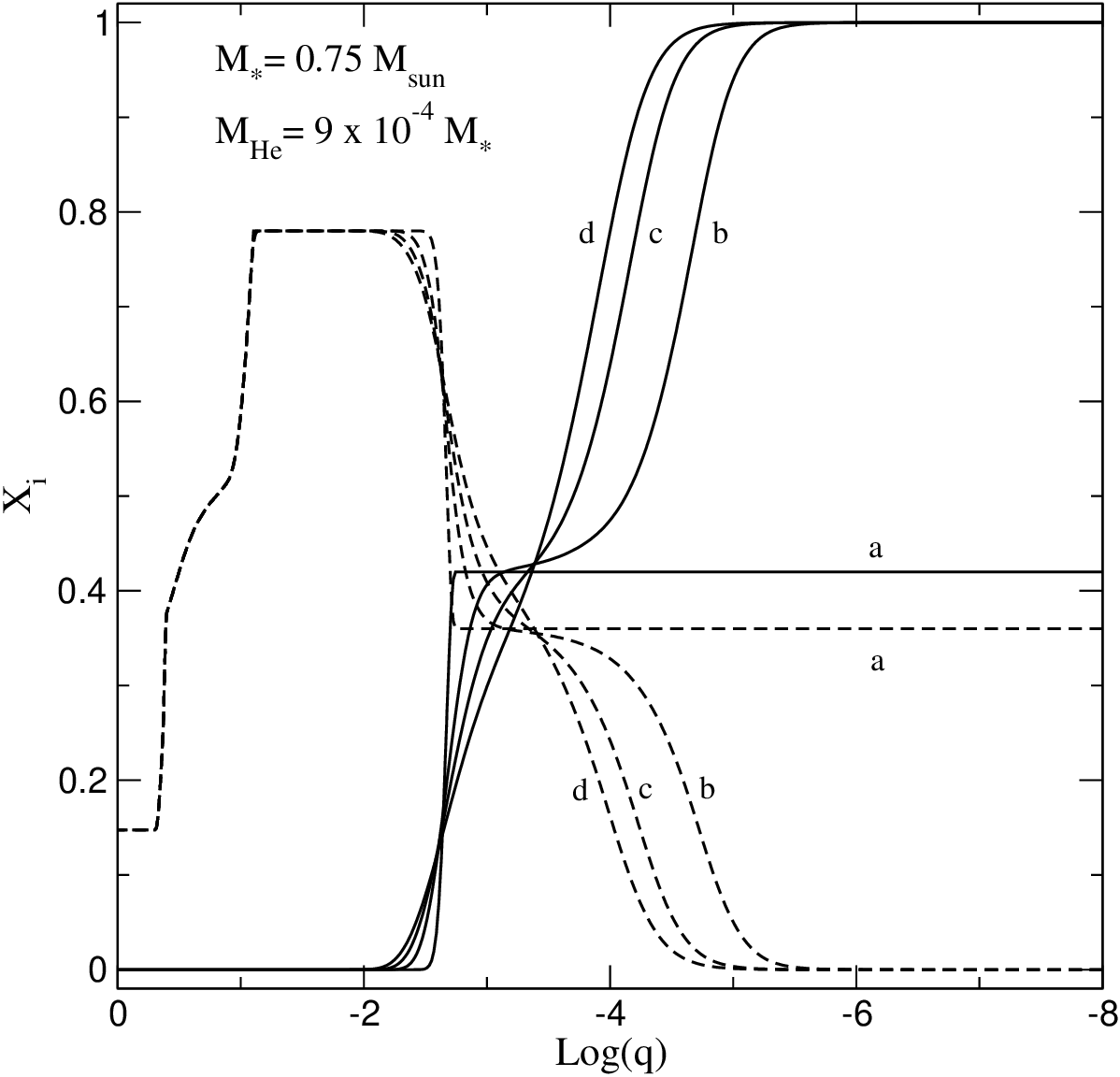}
\caption{Same as Fig. \ref{06_9d-4teff.eps} but for a 
stellar mass of 0.75 \msun. Starting from a model with an initially
homogeneous envelope (curves a), following models (b, c and d)
correspond to evolutionary
stages characterized by $T_{\rm eff}=$ 27200, 23100 and 19900 K. 
Note that, a single-layered profile emerges at larger \teff
values as compared with the 0.6 \msun model.} 
\label{075_9d-4teff.eps}
\end{figure}

\begin{figure}
\centering
\includegraphics[width=250pt]{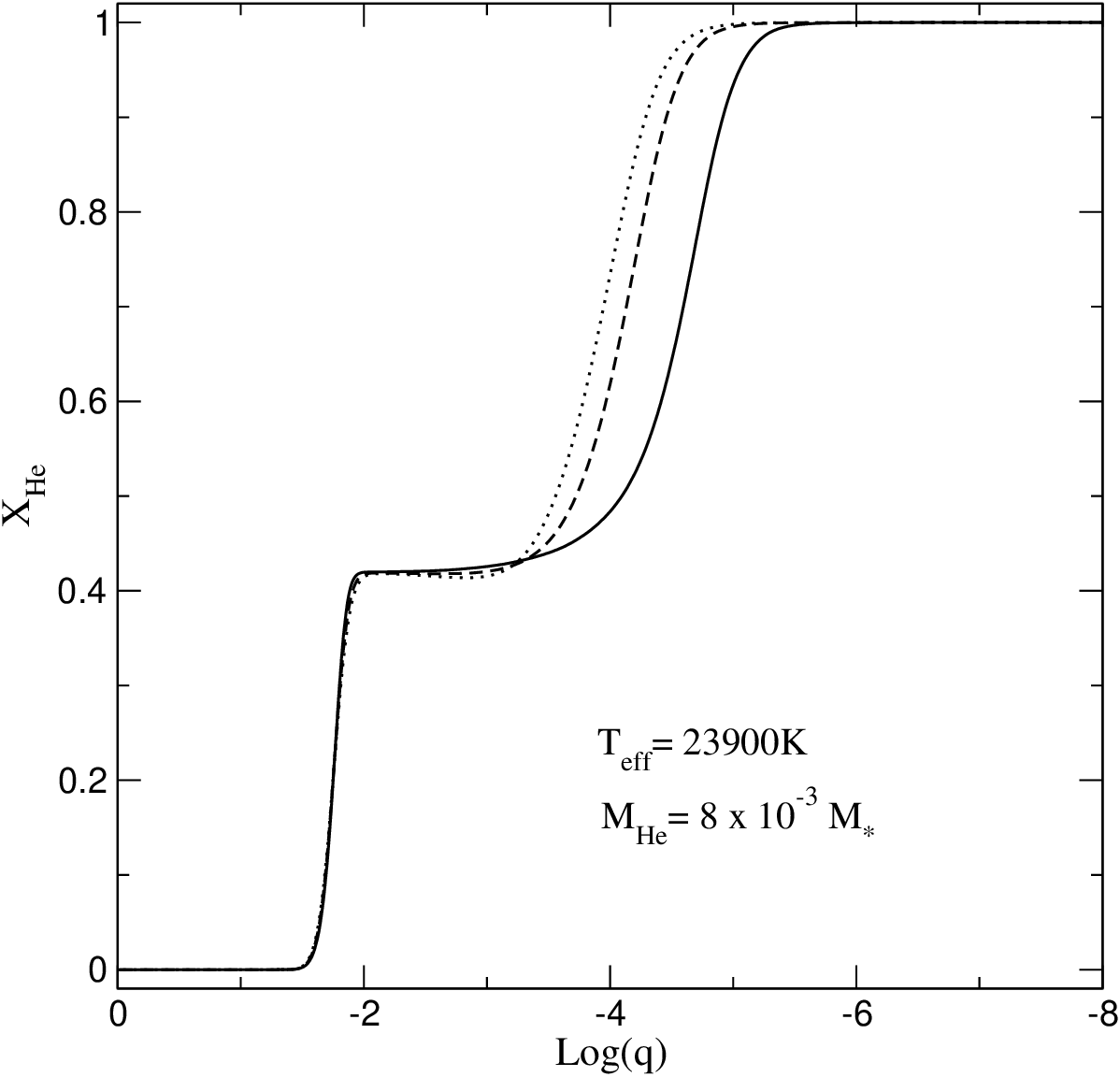}
\caption{
Abundance by mass of $^{4}$He 
as a function of the outer mass fraction 
($q$) for 0.60, 0.75 and 0.85-\msun\ DB models (solid, dashed and dotted
lines, respectively), having a helium content of $M_{\rm He}= 
8 \times  10^{-3} M_*$. Models correspond to \teff= 23900K.}
\label{difmasas1.eps}
\end{figure}

\begin{figure}
\centering
\includegraphics[width=250pt]{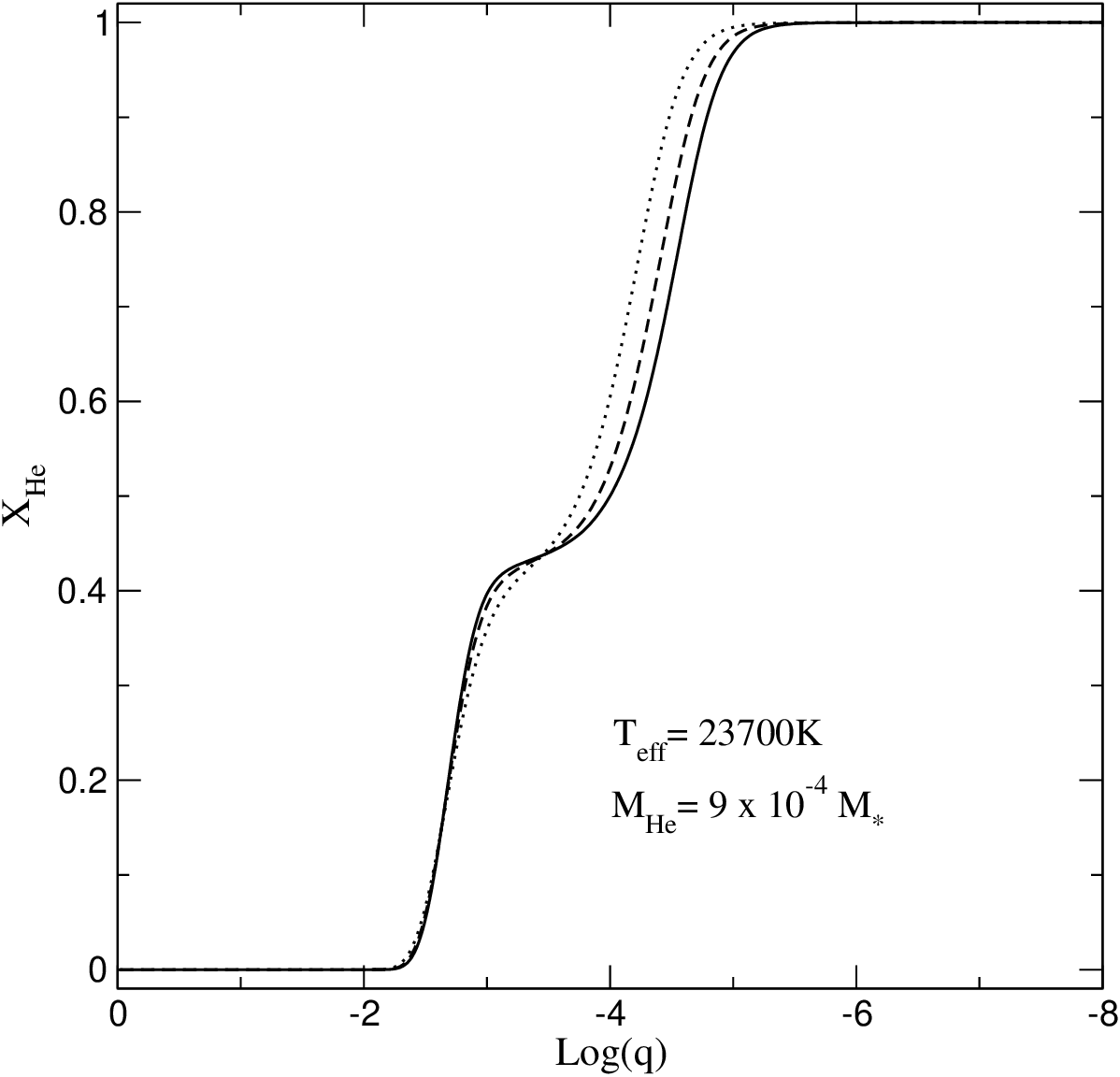}
\caption{Same as Fig. \ref{difmasas1.eps} but for a helium 
content of $M_{\rm He}= 9 \times  10^{-4} M_*$.
Solid, dashed and dotted lines correspond to DB models with
0.60, 0.70 and 0.75-\msun\ at \teff= 23700K.}
\label{difmasas2.eps}
\end{figure}

The dependence of the chemical profile evolution upon the stellar mass
is shown in Fig. \ref{075_9d-4teff.eps} to
\ref{difmasas2.eps}. In  particular, in Fig.  \ref{075_9d-4teff.eps} 
we compare  
the evolution of the chemical  profile for  0.75-\msun\ DB models  
with the  same helium content than that considered  in Fig.  
\ref {06_9d-4teff.eps} ($M_{\rm
He}= 9 \times  10^{-4} M_*$). The role of diffusion in relatively massive
DB white dwarfs is
clearly documented in this figure. Indeed, here,  over most  of the 
temperature range of  DB variables, the chemical
profile bears no signature of a double-layered structure. By contrast,
in the case of 0.6-\msun\  models the double-layered profile turns into
a single-layered  one only after evolution has  proceeded to effective
temperatures  somewhat below  the  red  edge  of  the  instability  strip.   In
Figs. \ref{difmasas1.eps} and \ref{difmasas2.eps} we compare the shape
of the  helium profile  for various stellar  masses at a  fixed \teff.
Note  that the  shape of  chemical profile  is very  dependent  on the
stellar mass of the star.

\begin{figure}
\centering
\includegraphics[width=250pt]{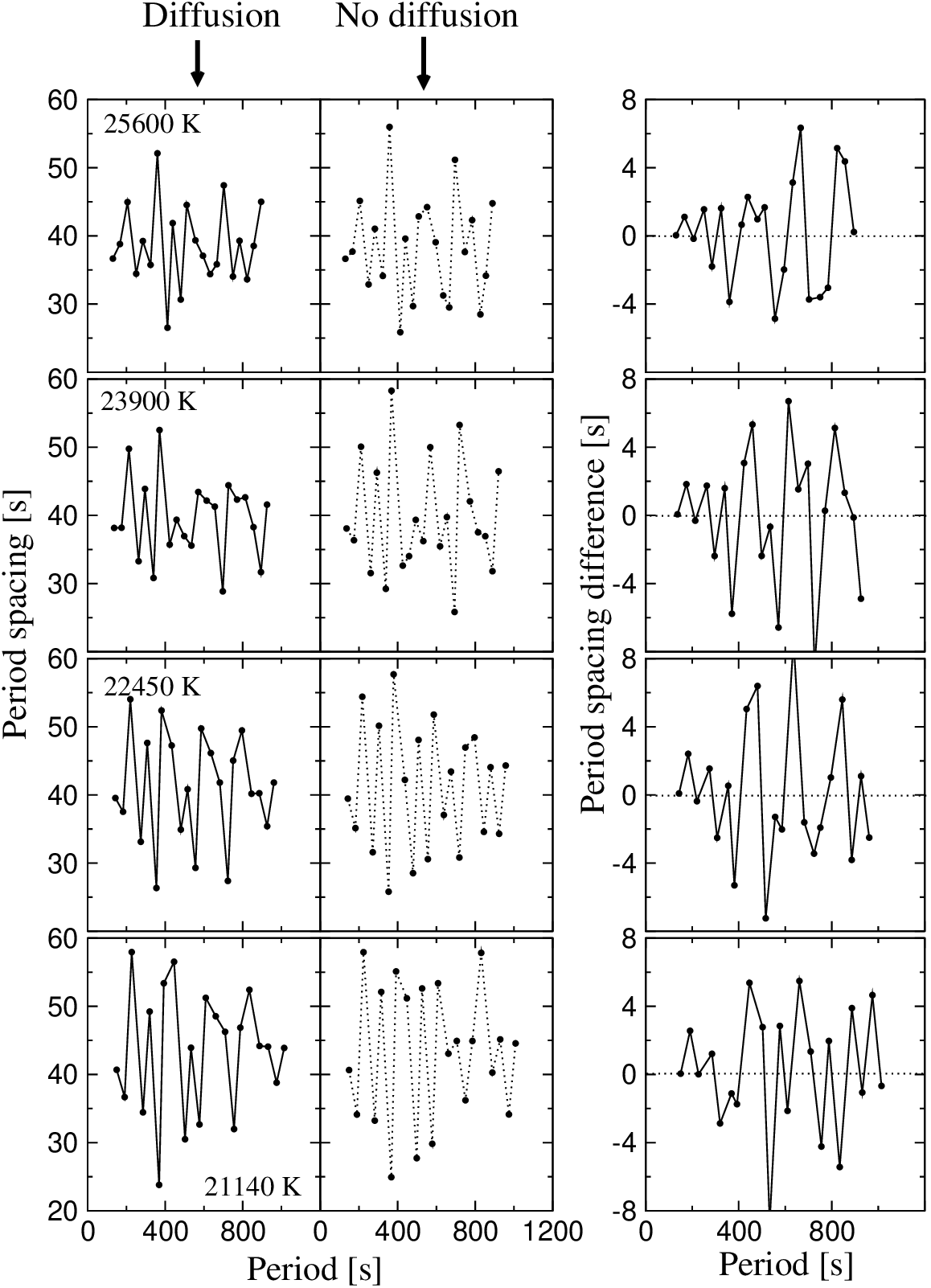}
\caption{Period spacing diagrams (for $\ell= 1$ modes) for the 
case of 0.6-\msun\ WD models with a helium 
content of $M_{\rm He}= 8 \times  10^{-3} M_*$. From top to bottom
panels correspond to decreasing effective temperatures (labels in the 
left panels). The left (centre) column of panels shows the results 
obtained when 
element diffusion has (not) been considered in the computations. Right 
panels depict the differences between the period spacing as given by 
diffusively 
evolving profiles and a fixed one predicted by diffusion at \teff 
$\approx$  27000K.}
\label{dp-8D-3.eps}
\end{figure}

\begin{figure}
\centering
\includegraphics[width=250pt]{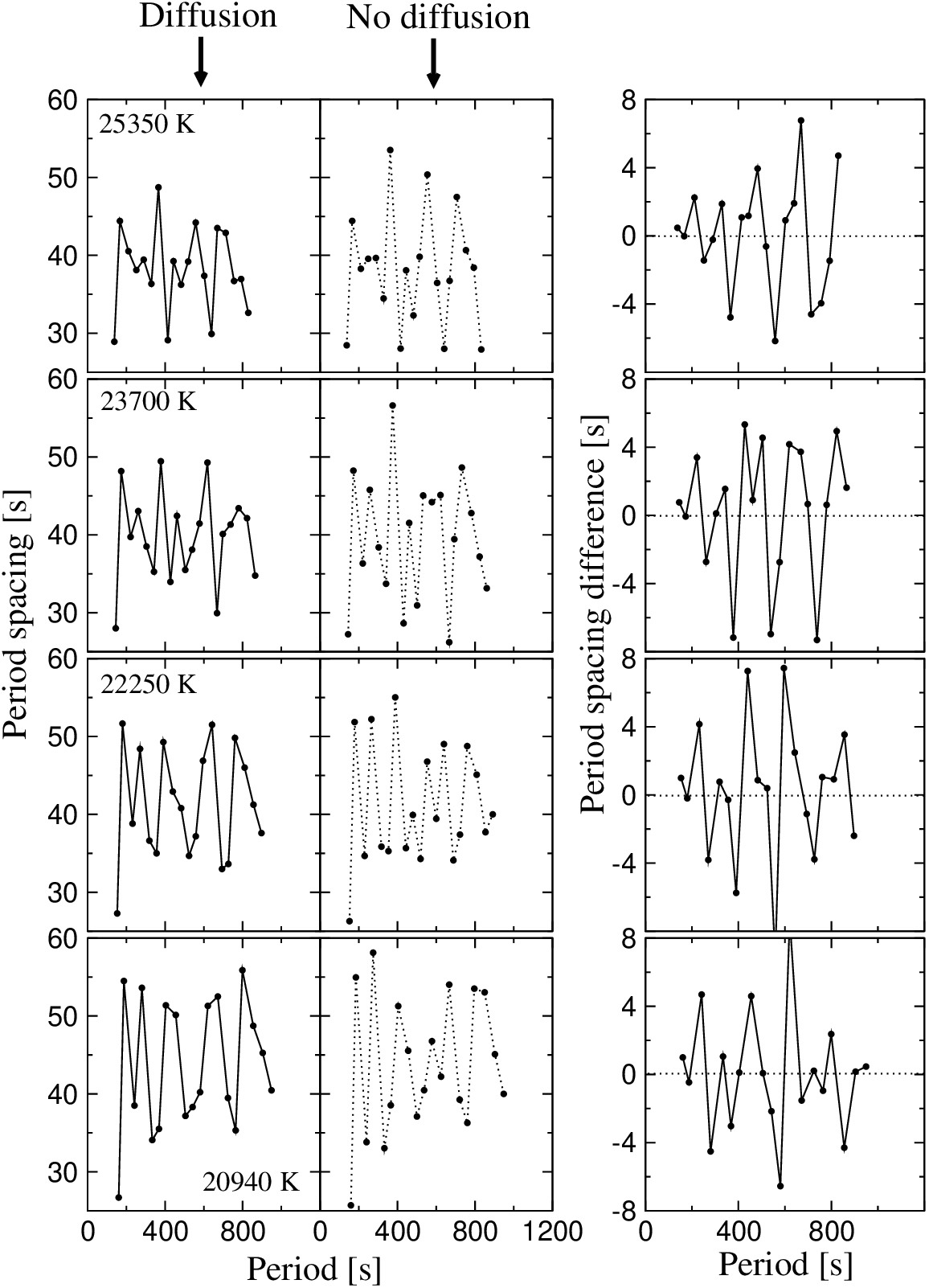}
\caption{Same as Fig. \ref{dp-8D-3.eps} but for a helium 
content of $M_{\rm He}= 9 \times  10^{-4} M_*$.}
\label{dp-9D-4.eps}
\end{figure}

\begin{figure}
\centering
\includegraphics[width=250pt]{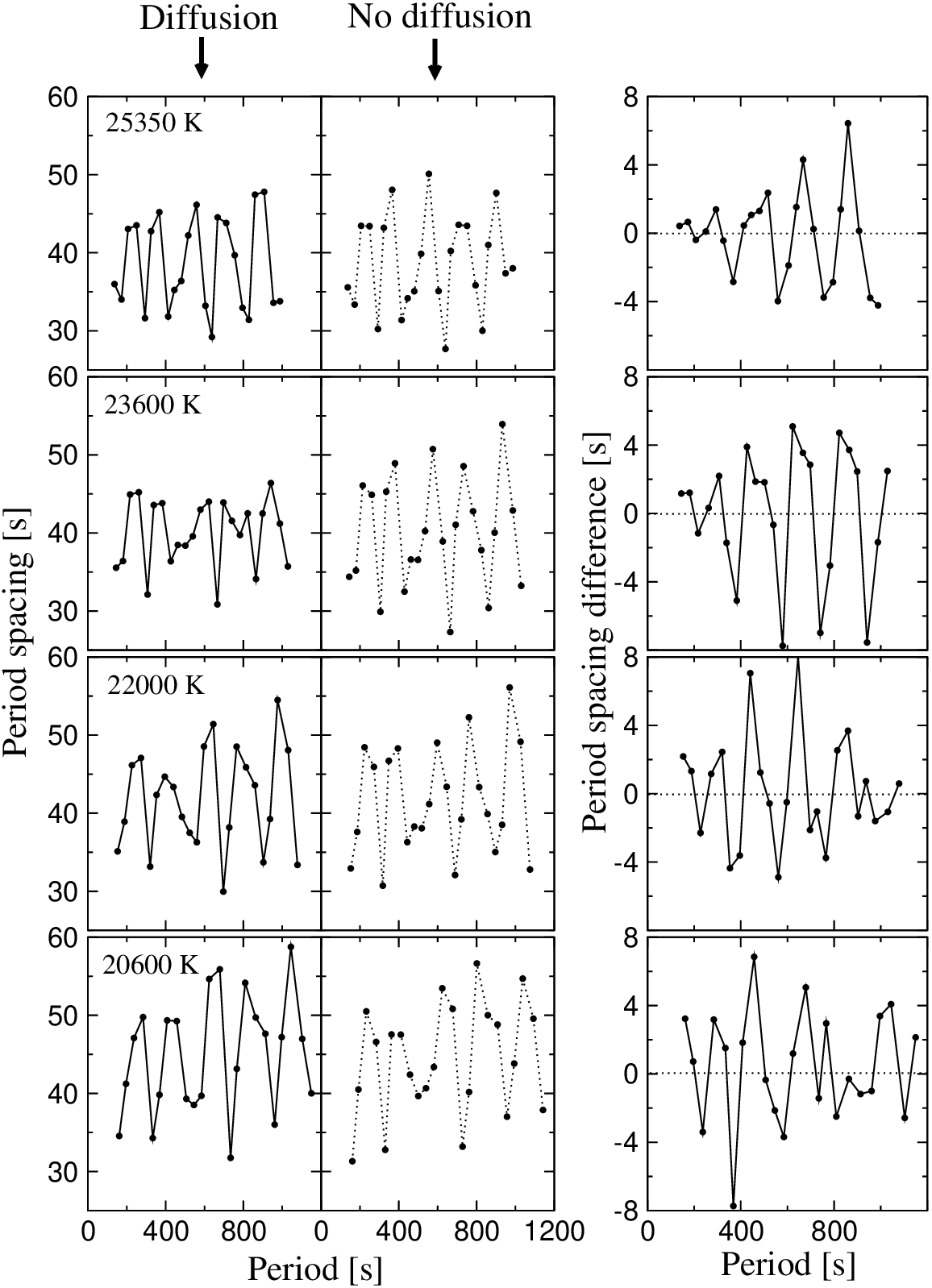}
\caption{Same as Fig. \ref{dp-8D-3.eps} but for a helium 
content of $M_{\rm He}= 1 \times  10^{-4} M_*$.}
\label{dp-1D-4.eps}
\end{figure}

Next, we  examine the  pulsational properties of  our   models.  
We shall consider non-radial ($\ell=1$) $g$-modes with periods $P_k$ ($k$
being  the  radial  overtone)  in  the  range  of  100-1000  s.  Figures
\ref{dp-8D-3.eps}, \ref{dp-9D-4.eps} and
\ref{dp-1D-4.eps} display the  forward period spacing ($\Delta  
P_k \equiv  P_{k+1} - P_k$) as a  function of period 
(period spacing diagrams) for  the case of
0.6-\msun\ DB white dwarf  models with helium contents of  
$M_{\rm He}= 8 \times 10^{-3}  M_*$, $9  \times 10^{-4}  M_*$  
and $1  \times 10^{-4}  M_*$,
respectively.  In   each  figure   results
corresponding  to a  sequence of  four selected  models  at decreasing
effective temperatures (as labeled in left panels) covering
the pulsational strip are depicted. The left column of panels (solid lines) 
depicts the $\Delta  P_k$ distributions as predicted by  calculations in which
element diffusion  has been accounted  for, and the central  column of
panels  (dashed  lines) shows  the  same information but  for  the  case 
in  which
diffusion processes have been locked in the evolutionary computations,
e.g.,  the  chemical  profiles  remain  fixed with further evolution  
from  a  \teff  value corresponding approximately to the blue edge  
of the DB instability strip. Finally, the right column of panels 
documents the differences between the values
of  $\Delta P_k$  as given  by diffusively  evolving profiles  and the
fixed ones.

An  inspection  of  the  figures   indicates  that  in  both  sets  of
computations the pulsation mode spectrum changes markedly as the white 
dwarf models
cool down  to the red  extreme of the  DB instability strip.  In the
case of models with diffusion, such changes are caused by:
(1) the  changes  the  mechanical and  thermal  structure 
of models in response to energy looses, changes which are
translated mainly into a decreasing in the Brunt-V\"ai\"sal\"a frequency
at the  core region  and the consequent  increasing of  periods and
period spacings, {\it and} (2)  the evolution of the chemical profiles
caused by diffusive processes,  thus  giving rise to 
changes in  the location of the well known  Brunt-V\"ais\"al\"a 
frequency's bumps associated with chemical interfaces. 
Instead, in  the case of  models in which chemical  diffusion is not 
considered, the  changes  in the  period
pattern are  due {\it only} to the secular changes  resulting from  
cooling (item 1), 
e.g., the Brunt-V\"ais\"al\"a  frequency decreases at  the innermost regions
but the profile of the Ledoux  term $B$ (a quantity closely involved 
in the computation of the Brunt-V\"ais\"a\"la frequency at  the chemical
interfaces;  see Brassard  et  al. 1991  for  details) remains  almost
fixed. Interestingly enough, during the instability strip the period 
spacing distribution 
is strongly modified according to  whether element diffusion is  
considered or  not.  

It is interesting to note that  the period spacing 
diagrams of the pulsation mode spectrum exhibits a clear mode-trapping 
substructure when a 
multi-layered structure is present, and more importantly how this 
substructure change due to diffusion. In fact, from left panels of Fig. 
\ref{dp-8D-3.eps} it can be seen that there is substructure at all 
evolutionary stages 
depicted, whereas in Fig. \ref{dp-9D-4.eps} the period spacing diagram 
becomes steadly simpler 
as the \teff decreases. Finally note that in Fig. \ref{dp-1D-4.eps} 
the period spacing diagram is characterized by a fairly clean structure at 
all temperatures shown. The 
occurrence of substructures in the period diagrams is of course 
intimately related to the presence or not of the double-layered structure, 
a fact  that can be related directly to Figs. \ref{06_7d-3teff.eps}, 
\ref{06_9d-4teff.eps} and \ref{06_1d-4teff.eps}. That is, the 
substructure  is determined by whether or not the 
pure helium layer has had time to engulf the whole helium-carbon-oxygen 
intershell. In the case of a single-layered structure there is a 
single strong mode-trapping cycle, otherwise there are two cycles of 
differing strength.

Finally, the period spacing differences between model with and 
without diffusion (as they were defined before)  
are by no means negligible and, as can be visualized from the plots, 
amounts to $\approx  \pm  8$ s  for  all  the helium  contents we  have
considered. These differences far exceed the quoted values of rms
differences  between  the  observed  and  calculated  period  spacings
($\sigma_{\Delta P}$) in  current detailed asteroseismological fits to
pulsating DB white dwarf stars. In light of these results, we conclude 
that the  dependence of the  shape  of  the  evolving  chemical  
profiles  on  the  effective temperature  cannot be  entirely ignored,  
that is, in order to maintain physical consistency, these parameters 
cannot be considered as independent ones when attempts at performing 
detailed seismological fit to a particular  star are made. These 
parameters are actually not coupled if the helium mass is less 
massive enough for a single-layered profile to emerge. 

\section{Conclusions and discussion}

In this paper we have explored the evolution of white dwarf stars with
helium-rich  atmospheres  (DB)  in  a  self-consistent  way  with  the
predictions of  time-dependent element diffusion. The focus  is on the
chemistry variations along the effective temperature range where these
stars   are  found   to   pulsate.   There   exist  theoretical   and
observational  evidence (Dehner  \&  Kawaler 1995;  Dreizler \&  Heber
1998) suggesting  that the  majority of  the DB
white  dwarfs  could be the  descendants  of PG1159  stars,  the  hot  and
hydrogen-deficient remnants of post-AGB  stars that experienced a very
late  helium thermal  pulse  on  their early  cooling  phase 
(see Herwig et al. 1999). In  this
regard, we have assumed for our starting DB white dwarf models
an uniform  envelope chemical composition  rich in helium,  carbon and
oxygen.  For this study, we have considered DB white dwarf models with
stellar masses of  0.60, 0.65, 0.70, 0.75 and 0.85  \msun. The mass of
the helium envelope ($M_{\rm He}$)  was varied in the range $10^{-2} -
10^{-4}  M_*$. For  a  consistent treatment  with  the predictions  of
element diffusion, OPAL  radiative opacities for arbitrary metallicity
including carbon-and oxygen-rich compositions have been employed.  Our
treatment  of  element  diffusion,  based on  the  multicomponent  gas
description of Burgers (1969), includes the effects of thermal diffusion,
which,  as we have shown,  lead to  DB models  with  thicker pure
helium mantles.

In   view  of   the   lack  of   evolutionary   calculations  with 
time-dependent  diffusion  for DB white dwarfs with different stellar 
masses and helium envelopes,  an initial  question  to  be addressed  
in  this work  was whether the double-layered chemical  profile 
expected for hot variable
DBs (assuming  that these  stars descend from  PG1159 stars)  could be
markedly modified  by further chemical  evolution before  the red
edge  of the  instability strip is  reached.  This  is  an important  point
because as  recently demonstrated by  Fontaine \& Brassard  (2002) the
presence of a double-layered structure leads to a distinct theoretical
period   distribution   from    that   predicted   by   single-layered
configurations.  Within the scope of the present paper is also
the study of the consequences of diffusively evolving  chemical
profiles for the adiabatic pulsational properties of DB white dwarfs.

Our  results  show  that,  as   found  in  previous  studies,  a
double-layered  envelope  is expected  by  the  time  the white  dwarf
reaches the  domain of the DB pulsational  instability. In particular,
element diffusion not only thickens  the pure helium mantle as cooling
proceeds but also it modifies the {\it shape} of the chemical profile,
an  aspect  to  be   considered  in  asteroseismological  fittings  of
pulsating  DB  white  dwarfs.  In   this  regard,  we  find  that  the
diffusively evolving shape of the envelope composition translates into
a  distinct  behaviour  of  the  theoretical  period  distribution  as
compared with the case in which the shape of the profile is assumed to
be fixed  during the DB  white dwarf evolution across  the instability
strip. In particular, we find that the presence of a double-layered 
structure causes the period spacing diagrams to exhibit 
mode-trapping substructures which change due to diffusion.

Another  finding of  this work  is  the fact  that the  double-layered
structure is  altered by further  chemical evolution to such  a degree
that, depending on the  helium content, a single-layered configuration
emerges even during the stages of pulsation instability. In particular,
for a  0.6 \msun\ DB  white dwarf with  $M_{\rm He}= 10^{-3}  M_*$ this
occurs  by the  time the  star reaches  the \teff  value  of $\approx$
20000K,  and   for  $M_{\rm   He}=  10^{-4}  M_*$,   a  single-layered
configuration  is  expected  at  \teff  $\approx$  27000K.   Thus,  if
post-AGB, DB white dwarf progenitors  are formed with a helium content
smaller than  $\approx 10^{-3}  M_*$, the double-layered  structure is
expected to become single-layered by the time the star reaches the red
edge of the DB instability strip. A  helium
envelope  mass of  $\approx 10^{-3}  M_*$  is more  appropriate for  a
massive white  dwarf. In particular, we  find that  the double-layered
profile in  a $\approx$ 0.8 \msun\ DB  turns into  a single-layered one
well before the red edge of the DB instability domain is reached.

The existence of 0.6 \msun\ DB  white dwarfs with a helium content less
massive than, say, $\approx 10^{-3}  M_*$ is difficult to reconcile with
the predictions of  the born-again scenario  for the formation
of   hydrogen-deficient    post-AGB   stars. Indeed, Herwig et al. (1999)  
find that after the end of the
late helium  thermal pulse the total  helium content left  in the star
amounts  to  $M_{\rm He}  \approx  10^{-2}  M_*$.   Now, there  exist
observational  evidence suggesting that  post-AGB mass-loss
episodes could  reduce (and even  remove) the mass of  the helium-rich
envelope considerably.  As recently  emphasized by Werner (2001) it is
possible that the surface chemistry  of all PG1159 stars is determined
not only by previous interior mixing  processes but also by mass-loss episodes.
Mass-loss rates in the  range $10^{-7}-10^{-8}$\msun/yr  are detectable 
in many luminous  PG1159  stars.   In  addition, tentative  evidence  for  the
persistence of mass-loss rates of the order $10^{-7}-10^{-10}$
\msun/yr down to the domain of hot helium-rich white dwarfs has been
presented (see  Werner 2001).  The  existence of such  mass-loss rates
would imply that  most  of the helium-rich envelope of DB progenitors
could  be  substantially reduced  during  the  $10^5-10^6$ yr  elapsed
during the post-AGB  stage. It is worth mentioning  that the existence
of PG1159  stars with a helium  content as low as  $10^{-3}$ \msun\ has
been suggested by asteroseismology in at least one of these stars with
a stellar mass  of 0.6 \msun (Kawaler \&  Bradley 1994), thus implying
the  occurrence of  modest mass-loss  during evolution  to  the PG1159
phase.  In light of such evidence, the possibility of the existence of some 
DB white dwarfs  with helium  contents lower  than  $10^{-3} M_*$ could 
not be discounted. According  to our calculations, a double-layered structure
would  thus not  be  expected for  such  white dwarfs  once they  have
reached the domain of DB pulsational instability.

Finally, we  have extended the scope of  our evolutionary calculations
down  to  effective temperatures  characteristics  of the  helium-rich
carbon-contaminated DQ  white dwarfs, the  supposed cooler descendants
of DBs.  The presence of carbon in the surface of such stars is widely
believed to  result from convective dredge-up of  the carbon diffusive
tail  by the  superficial  helium convection  zone  (Pelletier et  al.
1986, see also Koester et al. 1982).  New dredge-up carbon simulations
in DQ  stars by MacDonald  et al. (1998)  suggest that the  low carbon
abundance  observed in  many  DQs could  be  explained in  terms of  a
dredge-up effect only  if such white dwarfs are  characterized by very
thick helium  envelopes.  As our results  shown, if we  assume that DQ
white dwarfs may be linked to the post-AGB PG1159 stars via the 
PG1159-DB-DQ connection, then a
0.6 \msun\  DQ white dwarf  with a helium  content more massive  than 
$10^{-2} M_*$ would be  characterized by a double-layered structure. In
particular, we find that the pure  helium mantle extends down only  
to a fractional mass depth  $q \approx 5  \times10^{-5}M_*$ at 
\teff=  10000K, shallow enough for carbon to be  dredged-up to the 
surface with abundances far exceeding  the  observational  predictions.  
If  canonical  convective dredge-up  theory  is  correct,   our  results  
seem  to  discount  an
evolutionary  connection  between {\it  some}  DQs, those having low  carbon
surface   abundances,  and   the  PG1159   stars.    Full  evolutionary
calculations taking  into account  the evolutionary stages  leading to
the white dwarf formation, and particularly the chemical
structure emerging from the born-again scenario, are  required to place 
these assertions on a more solid basis. Work in this direction is in progress.

Detailed tabulations of the chemical profiles of our models
are available at http://www.fcaglp.unlp.edu.ar/evolgroup/

\begin{acknowledgements}

We warmly acknowledge Klaus Werner for sending us reprints
central to this work. It is a pleasure to acknowledge our referee,
whose comments and suggestions strongly improve the original version
of this papers.
This research was supported by the Instituto de Astrof\'{\i}sica La Plata.

\end{acknowledgements}


\begin{thebibliography}{}

\bibitem{} Althaus, L. G., Serenelli, A. M., C\'orsico, A. H., \&
Benvenuto, O. G. 2002, MNRAS, 330, 685

\bibitem{} Bradley, P. A., \&  Winget, D. E. 2001, ApJ, 430, 850          

\bibitem{}  Brassard, P.,  Fontaine,  G., Wesemael,  F.,  Kawaler, S.  D.,
\& Tassoul, M. 1991, ApJ, 367, 601

\bibitem{} Burgers, J. M. 1969, Flow Equations for Composite Gases
(New York: Academic)

\bibitem{} C\'orsico, A. H., Althaus, L. G., Benvenuto, O. G., \& Serenelli, A. M. 
2001, A\&A, 380, L17       

\bibitem{} Dehner, B. T., \& Kawaler, S. D. 1995, ApJ, 445, L141

\bibitem{} Dreizler, S., \& Heber, U. 1998, A\&A, 334, 618

\bibitem{} Fontaine, G., \&  Brassard, P. 2002, ApJ, 581, L33

\bibitem{} Gautschy, A., \& Althaus, L. G. 2002, A\&A, 382, 141           

\bibitem{} Heber, U. 1986, A\&A, 155, 33

\bibitem{} Herwig, F., Bl\"ocker, T., Langer, N., \& Driebe, T. 1999,
A\&A, 349, L5

\bibitem{} Iben, I. Jr., Kaler, J. B., Truran, J. W., \&  Renzini, A.
1983, ApJ, 264, 605 

\bibitem{} Iglesias, C. A., \& Rogers, F. 1996, ApJ, 464, 943

\bibitem{} Kawaler, S. D., \& Bradley, P. A. 1994, ApJ, 427, 415 

\bibitem{} Koester, D., Weidemann, V., \& Zeidler-K.T., E. M. 1982, A\&A, 
116, 147 

\bibitem{} MacDonald, J.,  Hernanz, M., \&  Jos\'e, J. 1998, MNRAS, 296, 523

\bibitem{} Magni, G., \& Mazzitelli, I.  1979, A\&A, 72, 134

\bibitem{} Metcalfe, T. S., Montgomery, M. H., \&  Kawaler, S. D. 2003, 
MNRAS, 344, L88 

\bibitem{} Metcalfe, T. S., Nather, R. E., \&  Winget, D. E. 2000, 
ApJ, 545, 974

\bibitem{} Metcalfe, T. S., Salaris, M. E., \& Winget, D. E. 2002, 
ApJ, 573, 803

\bibitem{} Metcalfe, T. S., Winget D. E., \& Charbonneau, P. 2001, ApJ, 557, 
1021

\bibitem{} Montgomery, M. H., Metcalfe, T. S., \& Winget D. E. 2001, ApJ, 548, 
L53

\bibitem{} Paquette,  C., Pelletier,  C., Fontaine,  G., \& Michaud, G.
1986, ApJS, 61, 177

\bibitem{} Pelletier,  C., Fontaine,  G., Wesemael, F., Michaud, G., \& 
Wegner, G. 1986, ApJ, 307, 242

\bibitem{} Salaris, M., Dom\'{\i}nguez, I., Garc\'{\i}a-Berro, E., 
Hernanz, M., Isern, J., Mochkovitch, R. 1997, ApJ, 486, 413

\bibitem{} Tassoul, M., Fontaine, G., \& Winget, D. E. 1990, ApJS, 72, 335

\bibitem{} Werner, K. 2001, Ap\&SS, 275, 27

\end{thebibliography}
\end{document}